\documentclass[12pt,aps,amsmath,latexsym,amsfonts]{JHEP3}
\usepackage[all]{xy}
\title{Ghosts in asymmetric brane gravity and the decoupled stealth limit}
\author{Kazuya Koyama$^1$, Antonio Padilla$^2$ and Fabio P Silva$^1$ \\
 $^1$Institute of Cosmology \& Gravitation,
University of Portsmouth,
Portsmouth~PO1~2EG, United Kingdom\\
$^2$School of Physics and Astronomy,
University of Nottingham,
Nottingham~NG7~2RD, United Kingdom}

\date{\today}

\abstract{We study the spectrum of gravitational perturbations around a vacuum de Sitter brane in a 5D asymmetric braneworld model, with induced curvature on the brane. This generalises the stealth acceleration model proposed by Charmousis, Gregory and Padilla (CGP) which realises the Cardassian cosmology in which power law cosmic acceleration can be driven by ordinary matter. Whenever the bulk has infinite volume we find that there is always a perturbative ghost propagating on the de Sitter brane, in contrast to the Minkowski brane case analysed by CGP.  We discuss the implication of this ghost for the stealth acceleration model, and identify a limiting case where the ghost decouples as the de Sitter curvature vanishes.}

\maketitle

\usepackage{epsfig}
\usepackage{amsfonts,amssymb,amsmath}
\usepackage{subfigure}

\newcommand{\mn}{\ensuremath{{\mu\nu}}}
\newcommand{\al}{\ensuremath{\alpha}}
\newcommand{\ga}{\ensuremath{\gamma}}

\newcommand{\La}{\ensuremath{\Lambda}}

\newcommand{\del}{\ensuremath{\partial}}

\newcommand{\half}{\frac{1}{2}}

\newcommand{\be}{\begin{equation}}
\newcommand{\ee}{\end{equation}}
\newcommand{\m}{\mathcal}
\newcommand{\ba}{\begin{eqnarray}}
\newcommand{\ea}{\end{eqnarray}}

\newcommand{\lab}[1]{{\label{#1}}}
\renewcommand{\tt}{\textrm}

\def\M{{\mathcal{M}}}

\begin{document}

\section{Introduction}
There is now a wealth of evidence \cite{sn, wmap, sdss} indicating that the universe is currently in a period of accelerated expansion.  One of the biggest challenges in cosmology today is understanding the origin of this late time acceleration.
One possibility is that $70\%$ of the energy content of the universe is
dominated by an as yet unknown form of energy, so-called dark energy. The most popular dark energy candidate is the vacuum energy, which takes the form of  a small and positive cosmological constant. In order to explain the
current acceleration, the value of the cosmological constant must contribute a vacuum energy density of the order $\rho_\La \sim 10^{-12} ({\tt eV})^4$. This is $10^{120}$ times smaller than what we might expect, given our current understanding of particle physics. Given that  particle physics is doing such a miserable job of explaining the accelerated expansion, it is important to look for alternative explanations.

A popular alternative is to interpret this acceleration as a sign that our understanding of gravity is breaking down, and that a large distance modification of Einstein's General Relativity is required.  Despite numerous attempts, it is fair to say that an established proposal has yet to  emerge that is consistent on both a fundamental and a phenomenological level. Arguably the most successful attempts have been inspired by the braneworld paradigm  (for a review see \cite{kkreview}).
In particular, the Dvali-Gabadadze-Porrati (DGP)  model~\cite{dgp} was discovered to have two cosmological branches, one of which gave rise to cosmic acceleration even when no matter was present on the brane \cite{dgpsa}.  This branch became known as the self-accelerating branch, for obvious reasons, but was later discovered to be  haunted by ghost instabilities around the vacuum de Sitter brane \cite{Luty, Nicolis,
arethereghosts, moreonghosts, unexorcised, bubbles, dgpspec, review, newperspective, saperts, modelforSA} (for a review see \cite{SAghosts}).  In this context a ghost is a field whose kinetic term has the "wrong" sign.  This pathology leads to a choice: either the ghost state has negative norm and unitarity is violated, or the ghost can have arbitrarily negative energy.  A ghost  in the perturbation  spectrum ({\it specter}oscopy!) indicates a catastrophic instability of the background, and therefore an unacceptably sick perturbative  theory. In DGP, the other cosmological branch (the "normal" branch),  is ghost-free but cannot be an alternative to $\Lambda$CDM since it still needs the introduction of the cosmological constant $\Lambda$ to explain the acceleration. Nevertheless it still has plenty of interesting
phenomenological features \cite{bwDE, arthur, phantoms}.

More recently, Charmousis, Gregory and Padilla (CGP)\cite{cgp}  presented a  generalisation of the DGP model in which they allowed for bulk curvature and introduced some asymmetry  across the brane \cite{battye, asymm1, asymm2, asymm_roy, asymm_kk, asymm_ggs}. This asymmetry could, in principle, apply to the bulk cosmological constant or even the bulk Planck scales, giving rise to a rich variety of cosmologies.  The authors focussed on those solutions that possessed asymptotically  Minkowski branes, despite the presence of self-accelerating solutions that they (correctly) assumed to be haunted by ghosts.  A subset of these solutions were shown to contain vacuum branes that were perturbatively stable, free from the ghoulish instabitities that terrorized the self accelerating DGP brane. The cosmological evolution of this subset was then analysed, and in some cases yielded extremely interesting results. Two limiting models in particular (the "decoupled" limit and the "conformal" limit)  were found to  exhibit power law acceleration but only when matter is present on
the brane. They dubbed this 'stealth acceleration'.

The cosmology is reminiscent of the Cardassian cosmology proposed by Freese and Lewis~\cite{cardassian}.  Here the standard Friedmann equation is modified so that $\rho \to \rho +c\rho^n$, where $n<2/3$, and one also finds that cosmic acceleration is driven by the presence of ordinary matter. The Cardassian model is an interesting empirical model, but did not have a concrete theoretical basis. The stealth model provides that by realising an effective Cardassian cosmology (with $n \approx 0.5$) within the braneworld paradigm.

In this paper we will consider vacuum de Sitter branes within the CGP set-up. This will include self-accelerating solutions, as well as the stealth models with some additional vacuum energy on the brane. We will study the spectrum of linearised perturbations about these solutions, closely following the corresponding analysis in the DGP model \cite{arethereghosts, moreonghosts, dgpspec}.  For an infinite volume bulk, we will find, without exception, that the vacuum is unstable because of the presence of ghosts. Just as for the self-accelerating branch of DGP, a ghost will manifest itself either through the radion mode, or through the helicity 0 mode of the lightest graviton. In some cases a ghost will also appear in the spin 1 sector.

The "decoupled" version of the stealth model is now of particular interest. We will find a class of de Sitter solutions that approach the "decoupled" model as the Hubble scale $H \to 0$. As the limit is approached the  ghost becomes more and more weakly coupled, until eventually it decouples completely.  We will infer some conclusions regarding the stability of the stealth models when matter is present.   For small $H$ it seems that we can carry our analysis of de Sitter branes over to the general Friedmann-Robertson-Walker case and conclude that the decoupled stealth model develops an instability albeit a very mild one softened by the weakness of the ghost coupling.  For larger $H$ the instability for de Sitter branes would be more severe, but it is not clear whether or not we can transfer this conclusion to the general FRW case.

The rest of this paper is organised as follows:  in section \ref{sec:setup} we describe the CGP model in detail, our generalisation, and the background solutions.  In section \ref{sec:perts} we analyse the spectrum of linearised perturbations and derive conditions for the presence of an helicity 0 ghost in the spin 2 sector. We study the coupling to matter in section \ref{sec:mat} and calculate the effective action in section \ref{sec:eff}. The effective action helps to reveal any further ghosts, including the radion ghost, which seems to take it in turns with the helicity 0 mode to haunt the background. We end with some concluding remarks in section \ref{sec:disc}.

\section{The CGP model:  set up and background solutions} \lab{sec:setup}
The CGP model is an asymmetric generalisation of its celebrated cousin, the DGP model.  In both models, our Universe is taken to be a 3-brane, $\Sigma$, embedded in between two five dimensional spacetimes, $\mathcal{M}_i$, where $i=L, R$.   In the original DGP scenario, we impose  $\mathbb{Z}_2$ symmetry across the brane, identifying $\M_L$ with $\M_R$ and having vanishing  vacuum energy in the bulk.  In the CGP model, however, we relax both of these assumptions.  The key new ingredient is the introduction of asymmetry.  Each spacetime $\M_i$ generically has a  five
dimensional Planck scale given by $M_i$, and a negative (or zero)
cosmological constant given by $\La_i=-6k_i^2$,. However, since we are no longer assuming $\mathbb{Z}_2$ symmetry across the brane, we can have  $M_L \neq M_R$  and $\La_L \neq \La_R$. Allowing for $\La_L \neq
\La_R$ is familar enough in domain wall scenarios \cite{DW}.  The Planck scale asymmetry is less familiar, but could arise
in a number of ways. Suppose, for example, that this scenario is
derived from a fundamental higher dimensional theory.  This theory
could contain a dilaton field that  is stabilised in different
fundamental vacua on either side of $\Sigma$. From the point of view
of a $5D$ effective description,  the $5D$ Planck scales would then
differ accordingly. Indeed naive expectations from string theory point
towards this asymmetric scenario as opposed to a symmetric one.
Different effective Planck scales can also appear on either side of a
domain wall that is bound to a five-dimensional  braneworld~\cite{nested}.

In keeping with the braneworld paradigm,  all matter and standard model interactions are confined to the brane, although gravity can propagate into  the fifth dimension. As in the DGP scenario, we include some intrinsic curvature induced on the brane.  This term is  rather natural and can be induced by matter loop
corrections~\cite{loops},  finite width effects~\cite{width} or even
classically from higher dimensional modifications of General
 relativity~\cite{z}. We will also include some vacuum energy on the brane in the form of some brane tension, $\sigma$.  At this point we introduce an important new development. In the original CGP paper,  the brane tension was fine-tuned against the bulk cosmological constants in order to  admit a Minkowski vacuum solution. This choice corresponds to having vanishing effective cosmological constant on the brane and was the analogue of the Randall-Sundrum fine-tuning.  In this paper we will introduce some additional tension so that the vacuum brane is de Sitter. Such detuning of brane tensions helped conjure up the ghost in the DGP model, and we will ultimately find that the same is true here.

This  set-up is described by the following action,
\begin{equation}
\lab{act}
S=\sum_{i=L,R} M_i^3\int_{\mathcal{M}_i}
\sqrt{-g}(R-2\Lambda_i)+2M_i^3\int_{\partial\mathcal{M}_i}\sqrt{-\gamma} K^{(i)} +\int_\Sigma\sqrt{-\gamma}(M_{4}^2
\mathcal{R}-\sigma+\m{L}_{ \tt {matter}}),
\end{equation}
where $g_{ab}$ is the bulk metric with corresponding Ricci tensor,
$R$.  The metric induced on the brane is  given by
$\gamma_{ab}=g_{ab}-n_an_b$
where $n^a$ is the unit normal to $\partial\mathcal{M}_i$ in
$\mathcal{M}_i$ pointing {\it out} of $\mathcal{M}_i$. Of course,
continuity of the metric at the brane requires that $\gamma_{ab}$ is
the same, whether it is calculated from the left, or from the right of
the brane. In contrast, the extrinsic curvature  of the brane can jump
from right to left.
In $\m{M}_i$,  it is defined as
\begin{equation} K^{(i)}_{ab}=\gamma^c_a
\gamma^d_b \nabla_{(c} n_{d)}, \lab{extrinsic}
\end{equation}
 with its trace  appearing in the Gibbons-Hawking boundary term in (\ref{act}). In the brane part of the action we have included the brane tension, $\sigma$,  and the  induced intrinsic
curvature  term, $\m{R}$, weighted by a $4D$  mass scale,
$M_4$. $ \m{L}_{ \tt {matter}}$ includes
any additional matter excitations.

The equations of motion in the bulk region, $\m{M}_i$,  are just the
 Einstein equations, with the appropriate cosmological constant, $\La_i$.
\begin{equation}
E_{ab}= R_{ab}-\frac{1}{2} R g_{ab}+\Lambda_i g_{ab}=0.
\lab{bulkeom}
\end{equation}
The equations of motion on the brane are described by the Israel
junction conditions, and can be obtained by varying the action
(\ref{act}), with respect to the brane metric, $\gamma_{ab}$.  This gives\footnote{The angled
brackets denote an averaged quantity at the brane. More precisely, for
some quantity $Q_i$ defined on the brane in $\partial \m{M}_i$, we
define the average
$\langle Q \rangle= \frac{Q_L+Q_R}{2}$.
Later on we will
also make use of the difference,  $\Delta Q =Q_L-Q_R$.}
\begin{equation} \Theta_{ab}=2\left \langle M^3 (K_{ab}-K
\gamma_{ab}) \right \rangle+M_{4}^2\left( \mathcal{R}_{ab}-\frac{1}{2}
\mathcal{R}\gamma_{ab}\right)+\frac{\sigma}{2}\gamma_{ab}=\frac{1}{2} T_{ab},
\lab{braneeom}
\end{equation}
where $T_{ab}=-\frac{2}{\sqrt{-\ga}}\frac{\partial
  \sqrt{-\ga}\m{L}_\tt{matter}}{\partial \ga^{ab}}$.
Note that the Israel equations here do not use the familiar
``difference'', because we have defined the unit normal as pointing
out of $\mathcal{M}_i$ on each side.  We adopt this (slightly) unconventional approach since it is more convenient in the asymmetric scenario where the brane is best thought of as the common boundary $\Sigma=\partial\M_L=\partial \M_R$.

We will now derive the vacuum solutions to the equations of motion
(\ref{bulkeom}) and (\ref{braneeom}). This corresponds to the case
where there are no matter excitations, and so, $T_{ab}=0$.  In each
region of the bulk, we introduce coordinates $x^a=(x^\mu, y)$, with
the brane located at $y=0$. We are interested in de Sitter brane solutions of the form
\begin{equation}
ds^2=\bar{g}_{ab} dx^adx^b =dy^2+N(y)^2 \bar \gamma_{\mu\nu}dx^\mu dx^\nu. \lab{background}
\end{equation}
where  $\bar \gamma_{\mu\nu}$ is the four dimensional de Sitter metric with curvature, $H$.  Inserting this into the bulk equations of motion (\ref{bulkeom}) gives
\begin{equation}
\left(\frac{N'}{N}\right)^2=\frac{H^2}{N^2}+k^2, \qquad \frac{N''}{N}=k^2, \lab{odes4a}
\end{equation}
where "prime" denotes differentiation with respect to $y$, and we have dropped the index $i$ for brevity. One can easily show that
\begin{equation}
	\lab{eq:aads}
	N(y) = \frac{H}{k}\sinh{k\,(y_h+\theta y)}, \quad y_h \equiv \frac{1}{k}\sinh^{-1} {k/H},
\end{equation}
where $\theta=\pm 1$.  Each region of the bulk corresponds to $0< y< y_\tt{max}$ where
\begin{equation}
y_\tt{max}=\begin{cases} \infty &  \textrm{for $\theta=1$}, \\
y_h  & \textrm{for $\theta=-1$} \lab{ymax}.
\end{cases}
\end{equation}
If we transformed to global coordinates in the bulk, $\theta=1$ would correspond to retaining the asymptotic region (large radius), whereas $\theta=-1$ would correspond to retaining the central region (small radius).  For $k \neq 0$, this means that when $\theta=1$ we keep the adS
boundary (growing warp factor) whereas when $\theta=-1$ we keep
the adS horizon (decaying warp factor).  Since we are interested in a modification of gravitational physics in the  infra-red, we will assume that the bulk volume is infinite,  and retain the asymptotic region on at least one side of the bulk. In other words, we do not consider the case  $\theta_L=\theta_R=-1$.

The boundary conditions at the brane (\ref{braneeom}) yield
\begin{equation}
6 \langle M^3 N'(0) \rangle+\frac{\sigma}{2}-3H^2 M_4^2=0, \lab{bc4a}
\end{equation}
so that the curvature $H$ is given by the real roots of
\be
\sigma=6M_4^2{H}^2-12\left\langle M^3 \theta
\sqrt{{H}^2+k^2}\right\rangle.
\ee
In  \cite{cgp}, the brane tension was fine tuned to a critical value, $\sigma_c=-6\langle M^3k\rangle$, so that the effective cosmological constant on the brane vanished. We now  introduce some additional tension $\epsilon>0$ so that $\sigma=\sigma_c+\epsilon$.  This introduces some positive curvature given by the roots of $\epsilon=F(H^2)$ where, as  in   \cite{cgp}, we have
\be
F(H^2)=6M_4^2{H}^2-12\left\langle M^3 \theta
\left(\sqrt{{H}^2+k^2}-k\right )\right\rangle \lab{F(H^2)}.
\ee
As in DGP, we have two classes of solution. There are those that vanish as $\epsilon \to 0$, so that we recover the Minkowksi brane studied in   \cite{cgp}, and there are those that approach a finite positive value, so that we have a de Sitter brane, even in the absence of an effective cosmological constant.   The former are the analogue of the normal branch in DGP, whereas the latter are the analogue of the self-accelerating  branch.  Of course, the class of solution depends on the form of the function $F(H^2)$, discussed in some detail in section 4 of \cite{cgp}. For example, the following represent  necessary and sufficient conditions for the existence of a normal branch solution:
\ba
 && M_4^2>\langle M^3 \theta /k \rangle,\\
 & \textrm{or~}& M_4^2=\langle M^3 \theta /k\rangle,  \langle M^3 \theta /k^3\rangle>0,\\
 & \tt{or~}& M_4^2=\langle M^3 \theta /k\rangle,  \langle M^3 \theta /k^3\rangle=0,  \langle M^3 \theta /k^5\rangle <0.
 \ea
Although we will study both classes of solution, we will be particularly interested in the normal branch since these will include small fluctuations about the  finely tuned "stealth" scenarios discussed in  \cite{cgp}.
\section{Vacuum fluctuations} \lab{sec:perts}
We shall now consider metric perturbations in the vacuum so that $g_{ab}=\bar g_{ab}+\delta g_{ab}$ and $T_\mn=0$.  In the unperturbed spacetime, given by (\ref{background}) and (\ref{bc4a}), the gauge was fixed in both $\m{M}_1$
and $\m{M}_2$ so that the brane was at $y=0$. However, a general
perturbation of the system must also allow the brane position to
flutter. In $\m{M}_i$, the brane will be located at
\be
y=\zeta_i(x^\mu).
\ee
It is convenient to work in a Gaussian Normal (GN) gauge, so that in $\M_i$ we have
\be
\delta g_{yy}=\delta g_{\mu y}=0, \qquad \delta g_\mn= h_{i \: \mn}(x, y).
\ee
In most of this discussion, we will drop the index $i$ although its should be understood that it is really there. Now, it is well known (see, for example, \cite{rs2}) that in the absence of any bulk matter, we may take $h_\mn$ to be transverse-tracefree $D^\mu h_\mn=h^\mu_\mu=0$. This is known as Randall-Sundrum gauge. It follows that the bulk equations of motion, $\delta E_{ab}=0$ give
\be
\left[\del_y^2+\frac{1}{N^2} (D^2-4H^2) -4k^2 \right]h_\mn(x, y)=0, \lab{bulkheqn}
\ee
where $D_\mu$ is the covariant derivative on the $4D$ de Sitter slicings,  and indices are raised/lowered using the $4D$ metric $\bar \ga_\mn$. To impose the boundary conditions at the brane, we need to apply a GN to GN gauge transformation that shifts the brane position back to $y=0$. The most general such transformation is given by
\be
y \to y-\zeta(x), \qquad x^\mu \to x^\mu-\xi^\mu(x)+D^\mu \zeta\int^y_0 \frac{dz}{N^2(z)},
\ee
so that
\be
h_\mn \to \bar h_\mn=h_\mn+h_\mn^{(\zeta)}+2N^2 D_{(\mu} \xi_{\nu)}. \lab{barh}
\ee
We call this new gauge "brane-GN" gauge. Although the brane position is fixed in this gauge, the original position $\zeta(x)$ still enters the dynamics through a bookkeeping term
\be
h_\mn^{(\zeta)}=-2\left(N^2\int^y_0 \frac{dz}{N^2}\right)D_\mu D_\nu \zeta+2NN' \bar \ga_\mn \zeta.
\ee
The metric perturbation in the new gauge is no longer transverse-tracefree, although it  is now straightforward to apply continuity of the metric at the brane
\be
\Delta \bar h_\mn (x, 0)=0, \lab{cont}
\ee
and the vacuum Israel equations (\ref{braneeom})
\be
\delta \Theta_\mn=-\left\langle M^3 \left(\frac{\bar h_\mn-\bar h \bar \ga_\mn}{N^2}\right)' \Big |_{y=0}\right\rangle +M_4^2 X_\mn(\bar h)=0, \lab{isrbarh}
\ee
where
\ba
X_{\mu \nu}(\bar h) &=&  \delta G_{\mu \nu} (\bar h)+ 3 H^2 \bar h_{\mu \nu} \nonumber \\
&=& -\half (D^2-2H^2) \bar h_\mn+D_{(\mu} D^\al \bar h_{\nu)\al}-\half D_\mu D_\nu \bar h \nonumber\\
&& \qquad \qquad \qquad  -\half \bar \ga_\mn\left[D^\al D^\beta \bar h_{\al\beta}- (D^2+H^2 )\bar h \right].
\ea
If we substitute the expression (\ref{barh}) into equation (\ref{isrbarh}) we find
\begin{eqnarray}
\left\langle M^3 \left(\frac{h_\mn}{N^2}\right)' \Big |_{y=0}+\frac{M_4^2}{2} (D^2-2H^2) h_\mn(x, 0)\right\rangle =\nonumber \\
2(D_\mu D_\nu-(D^2+3H^2) \bar \ga_\mn)\left\langle (M^3-M_4^2N'(0))\zeta\right \rangle.
\lab{isrh}
\end{eqnarray}
Note that this expression is independent of $\xi^\mu(x)$, as expected, since this just corresponds to diffeomorphism invariance along the brane. It is convenient  to decompose  $h_\mn$ in  terms of the irreducible representations of the $4D$ de Sitter diffeomorphism group
\be
h_\mn=h^{(2)}_\mn+h^{(1)}_\mn+h^{(0)}_\mn,
\ee
where $h^{(n)}_\mn$ corresponds to the spin $n$ contribution.  We can treat these modes independently of one another provided they have different masses\footnote{In $4D$ de Sitter, a transverse-tracefree tensor of mass $m$ satisfies $(D^2-2H^2)q^{(m)}_\mn=m^2 q^{(m)}_\mn$~\cite{ds}}. Let us now assume that this is indeed the case and analyse each spin separately. It will also be convenient to decompose the field $\xi_\mu(x)$ into its spin 1 and spin 0 components $\xi_\mu=\xi_\mu^{(1)}+\xi_\mu^{(0)}$. The field $\zeta(x)$ is just spin 0.
\subsection{Spin 2 modes} \lab{sec:spin2}
We begin by analysing the spin 2 modes. Since neither $\zeta$ nor $\xi_\mu$ have a spin 2 contribution, we can set them zero here, and  can further decompose the spin 2 piece of the metric by separating variables
\begin{equation}
	\lab{eq:modes}
	h_{\mu \nu}^{(2)}(x,y) = \int_m u_m(y) \chi^{(m)}_{\mu \nu}(x),
\end{equation}
where $\chi^{(m)}_\mn$ is a $4D$ tensor field of mass $m$ satisfying	$\left( D^2 - 2 H^2 \right)\chi^{(m)}_{\mu \nu}(x)=m^2 \chi^{(m)}_{\mu \nu}(x)$, and  $\int_m$ denotes a generalised sum, summing over discrete modes and integrating over continuum modes. The bulk equations of motion  (\ref{bulkheqn}) now give
\begin{equation}
	\lab{eq:bulkeom}
	u''_m(y) +\left(\frac{m^2  - 2H^2}{N^2} -4k^2 \right) u_m(y) = 0,
\end{equation}
This is easily solved in terms of the associated Legendre functions:
\begin{equation}
	\lab{eq:solads}
	u_m(y) =C_1 \left(\frac{k}{H}\right)^2 {\cal P}_{-1/2 \pm \nu}^{\pm2} \left(\coth{k (y_h+\theta y)}\right) + C_2 \left(\frac{k}{H}\right)^2 {\cal Q}_{-1/2 \pm \nu}^{\pm2} \left(\coth{k (y_h+\theta y)}\right),
\end{equation}
where $\nu = \sqrt{9/4 - m^2/H^2}$.    ${\cal P}^m_\nu(z)$ and ${\cal Q}^m_\nu(z)$ are the associated Legendre functions of the first and second kind, respectively. Of course, the expression (\ref{eq:solads}) is only well defined  for $m^2\leq \frac{9H^2}{4}$,   We could, in principle  analytically continue our solution  to  $m^2> \frac{9H^2}{4}$,  although this will not be necessary since our ultimate goal is to establish the existence of an helicity-0 ghost which is found in spin 2 modes of mass $0<m^2<2H^2$~\cite{spin2ghost}. Normalisability requires that~\cite{asymm_kk}
\be
\int_0^{y_\tt{max}} dy\, \frac{u_m^2}{N^2} < \infty,
\ee
so that for $\theta=1$ we only keep the part proportional to ${\cal P}^{-2}_{-1/2+\nu}(z)$, whereas for $\theta=-1$ we only keep the part proportional to ${\cal Q}^{2}_{-1/2+\nu}(z)$. Since we may assume that $u_m(0)=1$, without loss of generality, we get that the normalizable modes are given by
\be
	\lab{eq:modeads}
	u_m(y) = \begin{cases} \frac{{\cal P}_{-1/2 + \nu}^{-2} \left(\coth{k\,(y_h+y)}\right)}{{\cal P}_{-1/2 + \nu}^{-2} \left(\coth{k\,y_h}\right)} & \tt{for $\theta=+1$}, \\
	 \frac{{\cal Q}_{-1/2 + \nu}^{2} \left(\coth{k\,(y_h-y)}\right)}{{\cal Q}_{-1/2 + \nu}^{2} \left(\coth{k\,y_h}\right)}  &\tt{for $\theta=-1$}.  \end{cases}
\ee
 It will be instructive to take a closer look at two special cases. For massless modes, this expression simplifies to give
 \be
	\lab{eq:0mode}
	u_0(y) = \begin{cases} e^{-2ky}\left(\frac{2+\coth{k(y_h+ y)}}{2+\coth{k\,y_h}}\right) =\frac{N^2 \int_y^{y_\tt{max}} dz/N^4}{\int_0^{y_\tt{max}} dz/N^4}& \tt{for $\theta=+1$}, \\
	 N^2(y) &\tt{for $\theta=-1$}.  \end{cases}
\ee
whereas for "partially massless" modes of mass $m^2=2H^2$ we have
\be \lab{eq:usolssa}
	u_{\sqrt{2}H}(y)=\begin{cases} e^{-2ky} & \textrm{for $\theta=+1$},\\
	
	 \frac{NN'}{N'(0)} & \textrm{for $\theta=-1$}.
\end{cases}
\ee
Of course, neither the massless modes, nor the partially massless modes get excited in general. This is determined by the boundary conditions at the brane.  The spin 2 part of the continuity equation (\ref{cont}) now implies that $\Delta \chi^{(m)}_\mn(x)=0$ for each $m$, so that the spin 2 part of Israel equations (\ref{isrh}) yield the following quantization condition
\be
f(m^2)=
\left\langle M^3 \left(\frac{u_m}{N^2}\right)' \Big |_{y=0} \right\rangle+\frac{M_4^2}{2} m^2=0. \lab{fm^2}
\ee
Let us consider the lightest mode. For a finite volume bulk ($\theta_L=\theta_R=-1$), it is well known that this mode  is massless so that gravity looks four dimensional out to arbitrarily large distances. We do not consider this case here, and assume, without further loss of generality, that $\theta_R=+1$.  The lightest mode is now guaranteed to be  massive. If the mass lies in the forbidden region $0<m^2<2H^2$, then
this mode contains an helicity-0 ghost \cite{spin2ghost}. We can now check if such a mode exists, by application of Bolzano's theorem:
\begin{equation}
	\lab{eq:ghost}
	f(0) f(2H^2) < 0,
\end{equation}
since $f(m^2)$ is continuous over the forbidden region.  Although not {\it necessary} for the existence of a ghost, this condition is certainly {\it sufficient}. For an infinite bulk ($(\theta_L, \theta_R) \neq (-1, -1)$), it is easy enough to see that
\be
f(0)=-\frac{1}{2} \left \langle  M^3(1+\theta)\left[\int_0^{y_\tt{max}} \frac{dz}{N^4} \right]^{-1} \right\rangle <0.
\ee
This means we have an helicity-0 ghost whenever
\be
f(2H^2)=\left \langle \frac{ M^3}{2}\left(\frac{(1-\theta)H^2}{\sqrt{H^2+ k^2}}-2(1+\theta)( k+\sqrt{H^2+ k^2})\right)\right\rangle+ M_4^2H^2>0.
\ee
\subsection{Spin 1 modes}
We now turn our attention to the spin 1 modes, neglecting all contributions from spin 2 and spin 0.  Recall that  $\xi_\mu$ contains a spin 1 piece  $\xi_\mu^{(1)}(x)$, which is simply a divergence-free vector that can be  chosen in order  to guarantee continuity at the brane. The spin 1 part of the metric takes the form
\be
h_\mn^{(1)}=D_\mu A_\nu+D_\nu A_\mu,
\ee
where $A_\mu(x, y)$ is another divergence free vector. Since $h_\mn^{(1)}$ is transverse-tracefree, one can easily verify that $A_\mu$ behaves like a tachyonic vector  in $dS_4$, satisfying
\be (D^2+3H^2)A_\mu=0.
\ee
This tachyonic instability is a mild one, associated with the repulsive nature of inflating domain walls~\cite{ipser}. The metric contribution now resembles a massless spin 2 mode, $(D^2-2H^2)h_\mn^{(1)}=0$, and is therefore guaranteed not to mix with any of the genuine spin 2 modes discussed in the previous section. Furthermore, it follows that the profile in the bulk is given by the normalisable massless wavefunction (\ref{eq:0mode})
\be
A_\mu(x, y)=u_0(y)a_\mu(x).
\ee
The spin 1 part of the continuity equation (\ref{cont}), $\Delta (a_\mu+\xi_\mu^{(1)})=0$ , is trivially satisfied by choosing $\xi_\mu^{(1)}(x)=-a_\mu(x)$ on both sides of the brane. The Israel equations (\ref{isrh}) are independent of $\xi_\mu^{(1)}$, and require that
\be
\left\langle M^3\left(\frac{u_0}{N^2}\right)' \Big |_{y=0}a_\mu(x) \right\rangle=0. \lab{Abc}
\ee
If we assume, without any great justification, that $\Delta a_\mu=0$, if follows from (\ref{Abc}) that $f(0)a_\mu(x)=0$, and so $a_\mu(x)=0$. However, in a generalised asymmetric scenario there is no reason to assume that the spin 1 mode is symmetric.  More generally we can show that
\be
f(0)\langle a_\mu \rangle=\frac{1}{8} \Delta \left(  M^3(1+\theta)\left[\int_0^{y_\tt{max}} \frac{dz}{N^4} \right]^{-1}  \right)\Delta a_\mu,
\ee
which indicates that one spin 1 degree of freedom can, in principle, remain.

\subsection{Spin 0 modes}
We conclude this section with a study of the spin 0 modes, neglecting all contributions from higher spin. The brane bending piece $\zeta$ now plays a role, along with the spin 0 component of $\xi_\mu$, which takes the form $\xi_\mu^{(0)}=D_\mu \psi$, where $\psi(x)$ will be chosen in order to guarantee continuity at the brane. The spin 0 part of the metric perturbation can be written in terms of a pair of scalars, $\Phi(x,y)$ and $h^{(0)}(x, y)$, like so
\be
h_\mn^{(0)}=\left[D_\mu D_\nu-\frac{1}{4} D^2 \bar \ga_\mn\right] \Phi+\frac{1}{4} h^{(0)} \bar \ga_\mn.
\ee
From the transverse-tracefree property of $h^{(0)}_\mn$, it follows immediately that $h^{(0)}=0$ and
\be
(D^2+4H^2) \Phi=0.
\ee
Again, we have a mild tachyonic instability associated with inflating domain walls.  The metric contribution now resembles a "partially massless" spin 2 mode, $(D^2-2H^2)h^{(0)}_\mn=2H^2 h^{(0)}_\mn$, which could,  in principle,  mix with one of the genuine spin 2 modes discussed in section~\ref{sec:spin2}.  We will discuss this in more detail later on.   Assuming for the moment that there is no issue with mixing, we conclude that the scalar's  profile in the bulk is given by the partially massless wavefunction
\be
\Phi(x, y)=u_{\sqrt{2}H}(y) \phi(x).
\ee
The spin 0 part of the continuity equation is split into a pure gauge part, and a conformally de Sitter part. Requiring continuity of both parts separately implies that
\be
\Delta ( \phi+2\psi)=0, \qquad \Delta ( H^2 \phi+2N'(0)\zeta)=0. \lab{contphi}
\ee
The first condition can be trivially satisfied if we chose $\psi(x)=-\phi(x)/2$. The Israel equations (\ref{isrh}) are independent of $\psi$, and require that
\be
\left\langle\left( M^3\left(\frac{u_{\sqrt{2}H}}{N^2}\right)' \Big |_{y=0} +M_4^2 H^2\right) \phi(x)\right\rangle=2\langle (M^3-M^2_4 N'(0))\zeta \rangle. \lab{isrphi}
\ee
It follows from (\ref{contphi}) and (\ref{isrphi}) that
\be
\Delta \phi=-\frac{2}{H^2} \Delta\left [ \theta \zeta \sqrt{H^2+ k^2}\right ], \qquad \langle \phi \rangle=\alpha \left \langle \theta \zeta \sqrt{H^2+ k^2} \right\rangle +\beta \Delta \left[\theta \zeta \sqrt{H^2+ k^2} \right],
\ee
where
\ba
\alpha &=&\frac{2}{f(2H^2)}\left[ \left\langle  \frac{ M^3 \theta}{\sqrt{H^2+ k^2}} \right \rangle- M^2_4 \right], \\
\beta &=& -\frac{1}{4H^2f(2H^2)}\Delta\left[   M^3 (1+\theta) \left( \frac{(k+\sqrt{H^2+k^2})^2}{\sqrt{H^2+ k^2}} \right)\right].
\ea
Here we see that the fluctuation in the brane position sources the bulk mode $\phi(x)$. We therefore associate it  with the radion.  Again, there is no reason to assume $\Delta \phi=0$, so that in general the boundary conditions leave us with up to two spin 0 degrees of freedom.  Note that both $\alpha$ and $\beta$ diverge as $f(2H^2) \to 0$. This singular limit corresponds to the case where there exists a genuine spin 2 mode with mass $m^2=2H^2$. The divergence in $\alpha$ and $\beta$ reflects the fact that the lightest spin 2 mode is no longer orthogonal to the spin 0 contribution, and cannot be treated independently. The two modes mix and a more careful analysis is required. Finally, we also note  that  we can write  $\alpha=-F'(H^2)/3f(2H^2)$, where $F(H^2)$ is given by equation (\ref{F(H^2)}).

\section{Coupling to matter} \lab{sec:mat}
When we introduce some additional energy-momentum, $T_\mn$, on the brane, the homogeneous solution discussed in the previous section picks up an additional contribution that describes the responses of fields to the source on the brane,
\be
h_\mn(x, y) \to h_\mn(x, y)+ \pi_\mn(x, y),  \qquad \zeta_i(x) \to \zeta_i(x)+\pi_i(x),
\ee
where $\pi_\mn$ is transverse-tracefree. In analogy with the theory of ordinary differential equations, it is useful to think of the homogeneous pieces, $ h_\mn(x, y)$ and $\zeta_i(x)$, as the "complementary functions" and the inhomogeneous pieces, $\pi_\mn(x, y)$ and $\pi_i(x)$, as the "particular integrals". The "particular integrals"  must be solutions to the following
 \be
\left[\del_y^2+\frac{1}{N^2} (D^2-4H^2) -4k^2 \right]\pi_\mn(x, y) =0, \lab{pibulk}
\ee
\be
\Delta \left[ \pi_\mn(x, 0)+2N'(0) \bar \ga_\mn \pi(x) \right]=0,  \lab{picont}
\ee
\begin{gather}
\left\langle M^3 \left(\frac{\pi_\mn}{N^2}\right)' \Big |_{y=0}+\frac{M_4^2}{2} (D^2-2H^2) \pi_\mn(x, 0)\right\rangle =\nonumber \\ \hspace{3cm}2(D_\mu D_\nu-(D^2+3H^2) \bar \ga_\mn)\left\langle (M^3-M_4^2N'(0))\pi(x)\right \rangle
+ \half T_\mn. \lab{piisr}
\end{gather}
Tracing the two boundary conditions, and carrying out a little algebra, gives
\be
\Delta \left [N'(0) \pi(x)\right]=0,  \qquad  (D^2+4H^2)\left\langle N'(0) \pi(x)\right \rangle =-\frac{T}{2F'(H^2)}, \lab{mattercoupling}
 \ee
 where $F(H^2)$ is given by equation (\ref{F(H^2)}). This completely specifies the $\pi_i(x)$, since any homogeneous brane bending is already accounted for in the $\zeta_i(x)$. We now turn our attention to the $\pi_\mn$. The traceless part of (\ref{picont}) demonstrates that $\Delta \pi_\mn(x, 0)=0$, whereas the Israel equation (\ref{piisr}) may be rewritten like so
 \be
\left\langle M^3 \left(\frac{\pi_\mn}{N^2}\right)' \Big |_{y=0}+\frac{M_4^2}{2} (D^2-2H^2) \pi_\mn(x, 0)\right\rangle =\frac{1}{2} \tau_\mn(x),
\ee
where $\tau_\mn$ is a gauge invariant brane stress energy perturbation defined as~\cite{dgpspec}
\ba
\tau_\mn(x)&=&T_\mn-\frac{2}{3}F'(H^2)(D_\mu D_\nu-(D^2+3H^2) \bar \ga_\mn)\left\langle N'(0) \pi(x)\right \rangle \\
&=& T_\mn+\frac{1}{3}(D_\mu D_\nu-(D^2+3H^2) \bar \ga_\mn)\left(\frac{T}{D^2+4H^2}\right).
\ea
It turns out that in ${\cal M}_i$,
\be
\pi^{(i)} _\mn(x, y)=\int d^4 x' \sqrt{-\bar \ga} ~ G^{(i)}_\mn{}^{\al \beta}(x, y; x', 0) \tau_{\alpha \beta}(x'),
\ee
where $G^{(i)}_\mn{}^{\al \beta}(x, y; x', 0)$ is the relevant Green's function, satisfying
\be
\left[\del_y^2+\frac{1}{N^2} (D^2-4H^2) -4k^2 \right]G^{(i)}_\mn{}^{\al \beta}(x, y; x', 0) =0,  \lab{Gbulk}
\ee
\be
\Delta \left[ G _\mn{}^{\al \beta}(x, 0; x', 0)  \right]=0.  \lab{Gcont}
\ee
\begin{gather}
\left\langle M^3 \left(\frac{G_\mn{}^{\al \beta}(x, y; x', 0) }{N^2}\right)' \Big |_{y=0}+\frac{M_4^2}{2} (D^2-2H^2) G_\mn{}^{\al \beta}(x, 0; x', 0) \right\rangle =\frac{\delta^{(4)}(x-x')}{\sqrt{-\bar \ga}}. \lab{Gisr}
\end{gather}
The Green's function can be expressed in terms of the wavefunctions $u_m(y)$ discussed in section (\ref{sec:spin2}). Defining the {\it normalised} wavefunctions $ \hat u^{(i)}_m(y)={\cal N}_m u^{(i)}_m(y) $, where the ${ \cal N}_m$ are chosen so that
\be
\left \langle M_4^2 \hat u_m(0)\hat u_n(0)+ 2M^3 \int_0^{y_\tt{max}} dy ~\frac{\hat u_m \hat u_n}{N^2} \right\rangle=\begin{cases} \delta_{mn} & \tt{for discrete modes}, \\
\delta(m-n) & \tt{for continuum modes}, \end{cases}
\ee
we have
\be
G^{(i)}_\mn{}^{\al \beta}(x, y; x', 0)= -\int_p  \chi_\mn^{(p)}(x) \chi^{*(p)\alpha \beta}(x') \int_m  \frac{\hat u^{(i)}_m(y) \hat u^{(i)}_m(0)}{p^2-m^2}.
\ee
Note that  $(D^2-2H^2) \chi_\mn^{(p)}=p^2 \chi_\mn^{(p)}$, and $\chi^{* \alpha \beta}$ satisfies
$$\int_p \chi_\mn^{(p)}(x) \chi^{*(p)\alpha \beta}(x') =\delta^\al_\mu \delta^\beta_\nu \delta^{(4)}(x-x')/\sqrt{-\bar \ga}.$$
For more details on  this construction, at least for DGP gravity, see section 3.3 of~\cite{dgpspec}.

\section{The effective action} \lab{sec:eff}
We now compute the effective $4D$ action of normalisable vacuum perturbations. This will enable us to identify any ghosts: pathological modes with negative kinetic terms.  Of course, we already know that whenever $f(2H^2)>0$ a ghost haunts the helicity-0 sector of the lightest spin 2 mode. Our effective action calculation will reveal a generic spin-0 "radion"  ghost in the opposite regime, ie when $f(2H^2)<0$.

We begin our calculation in  bulk Randall-Sundrum gauge, so that the brane is positioned at $y=\zeta(x)$ and  the metric perturbation is given by
\be
h_\mn(x, y)=\int_m u_m(y)\chi^{(m)}_\mn(x)+u_0(y)h_\mn^{(a)}(x)+u_{\sqrt{2}H}(y)h_\mn^{(\phi)}(x),
\ee
where
\be
h_\mn^{(a)}(x) = D_\mu a_\nu +D_\nu a_\mu, \qquad h_\mn^{(\phi)}(x) =  (D_\mu D_\nu +H^2 \bar \ga_\mn) \phi.
\ee
In computing the action, it is  important  to leave the $4D$ fields off-shell. In other words, we do not assume $(D^2-2H^2)\chi^{(m)}_\mn=m^2\chi_\mn^{(m)}$, $(D^2+3H^2)a_\mu=0$, or $(D^2+4H^2)\phi=0$.  These equations should follow from variation of the action at the end of the calculation.

Randall-Sundrum gauge is the correct gauge choice far from the brane, since it contains no pure gauge modes with a non-normalisable profile in the bulk.  However, in order to compute the effective action, it is convenient to be in brane-GN gauge close to the brane so that it lies at $y=0$ and the $4D$ coordinates match on either side. This can be achieved whilst maintaining Randall-Sundrum gauge far from the brane, but only at a price: we are no longer everywhere Gaussian-Normal.  We can transform to this "fixed wall" gauge from everywhere Randall-Sundrum gauge by the following gauge transformation
\be
y \to y-\eta^y(x, y), \qquad  x^\mu \to x^\mu-\eta^\mu(x, y),
\ee
where
\be
\eta^y(x, y)=\begin{cases}
\zeta(x) & \textrm{for $y \ll y_*$}, \\
0 & \textrm{for $y \gg y_*$},
\end{cases} \qquad
 \eta^\mu(x, y)=\begin{cases}
\xi^\mu(x)-D^\mu \zeta(x) \int_0^y \frac{dz}{N^2(z)} & \textrm{for $y \ll y_*$}, \\
0 & \textrm{for $y \gg y_*$},
\end{cases}
\ee
where $0<y_*<y_\tt{max}$ is some appropriately chosen finite distance.  It follows that
\be
\delta g_{ab} \to \delta g_{ab}+2\nabla_{(a} \eta_{b)} \lab{dg},
\ee
where $\nabla$ is the covariant derivative for $\bar g_{ab}$. This new gauge interpolates between Randall-Sundrum gauge deep inside the bulk and brane-GN gauge near the brane. As result, the metric perturbation along the brane is the same as in brane-GN gauge, with
\be
\delta \ga_\mn=h_\mn(x, 0)+2N'(0)\bar \ga_\mn\zeta +2D_{(\mu} \xi_{\nu)}. \lab{dga}
\ee
We now perturb the action to quadratic order
\be
\delta S=\left \langle M^3 \int_{\mathcal{M}} d^5 x \sqrt{-\bar g} \delta g^{ab} \delta E_{ab}\right \rangle +\half \int_\Sigma d^4x \sqrt{-\bar \ga} \delta \ga^\mn \delta \Theta_\mn,
\ee
where $\delta E_{ab}$  and $\delta \Theta_\mn$ are the linearised  bulk equation of motion (\ref{bulkeom}) and  vacuum Israel equation (\ref{braneeom}), respectively. Using (\ref{dg}), (\ref{dga}) and the Bianchi identity $\nabla^a\delta E_{ab}=0$, we find that
\be
\delta S=\int d^4 x \sqrt{-\bar \ga} \delta \mathcal{L},
\ee
where
\begin{gather}
\delta {\cal  L}=\left\langle -M^3\left[ \int_0^{y_\tt{max}} dy \;h^\mn(x, y) \delta E_\mn(h)\right]+2M^3 \eta^a(x, 0) \delta E_{ay}(h)\Big |_{y=0} \right\rangle\nonumber\\
\qquad\qquad-\half \left\langle h^\mn (x, 0)+2N'(0)\bar \ga^\mn\zeta +2D^{(\mu} \xi^{\nu)}\right\rangle\delta \Theta_\mn.
\end{gather}
We cannot assume $D^\mu h_\mn^{(a)} =D^\mu h_\mn^{(\phi)} =h^{(\phi)}=0$, since these imply the on-shell equations of motion for $a_\mu$ and $\phi$. We therefore need the following expressions for $\delta E_{ab}$ and $\delta \Theta_\mn$ for a generic GN perturbation.
\ba
\delta E_\mn(h) &=& \frac{1}{N^2} X_\mn(h)-\half \left[ \del_y^2-2\left(\frac{H^2}{N^2}+2k^2\right)\right]\left(h_\mn-h\bar \ga_\mn \right), \\
\delta E_{\mu y} (h)&=& \half \del_y \left[\frac{D^\nu(h_\mn-h \bar \ga_\mn)}{N^2}\right], \\
\delta E_{yy}(h)&=&\frac{3 N'}{2N}\del_y\left[\frac{h}{N^2}\right]-\frac{1}{2N^4}(D^{\mu} D^\nu-(D^2+3H^2)\bar \ga^\mn)h_\mn, \\
\delta \Theta_\mn &=& -\left\langle \left[M^3 \del_y \left(\frac{ h_\mn- h \bar \ga_\mn}{N^2}\right)-M_4^2 X_\mn( h)\right]\Big |_{y=0} \right\rangle \nonumber \\
&& \qquad +2(D_\mu D_\nu-(D^2+3H^2) \bar \ga_\mn)\left\langle (M^3-M_4^2N'(0))\zeta\right \rangle.
\ea
Making use of equations  (\ref{eq:bulkeom}), (\ref{fm^2}),  (\ref{Abc}),   (\ref{contphi}), (\ref{isrphi}), as well as  the orthogonality condition
\be
\left \langle
2 M^3\int_0^{y_\tt{max}} dy \frac{u_m(y)u_n(y)}{N^2(y)}+M_4^2u_m(0)u_n(0)
\right \rangle
=0, \qquad m \neq n,
\ee
we arrive at the following $4D$ effective Lagrangian
\be
\delta {\cal L}=\delta {\cal L}_2+\delta {\cal L}_1+\delta {\cal L}_0,
\ee
where the spin 2, spin 1 and spin 0 contributions are respectively given by
\ba
\delta {\cal L}_2&=&\half \int_m \left[\int_0^{y_\tt{max}} dy \frac{u_m(y)^2}{N^2(y)}+M_4^2\right]\chi^{(m) \mn} (D^2-2H^2-m^2)\chi^{(m)}_\mn, \\
\delta {\cal L}_1 &=& \frac{1}{4}\left\langle \frac{1}{M^3(u_0/N^2)'|_{y=0}} \right\rangle^{-1}
\Delta a^\mu (D^2+3H^2) \Delta a_\mu, \\
\delta {\cal L}_0 &=& \frac{9  f(2H^2)}{F'(H^2)}\langle\gamma \rangle \left[\langle \phi\rangle+\frac{\Delta \ga\Delta \phi}{4\langle \ga \rangle}\right](D^2+4H^2) \left[\langle \phi\rangle+\frac{\Delta \ga\Delta \phi}{4\langle \ga \rangle}\right]   \nonumber \\
&& \qquad +\frac{3H^2}{8} \left\langle \frac{1}{\ga}\right \rangle^{-1}\Delta \phi (D^2+4H^2) \Delta \phi,
\ea
and
\ba
\gamma &=& M^3 \left(\frac{1+\theta}{2}\right)\frac{ (k+\sqrt{H^2+k^2})^2}{\sqrt{H^2+k^2}}>0, \\
(u_0/N^2)'|_{y=0} &=& - \left(\frac{1+\theta}{2}\right)\left(\int_0^{y_\tt{max}} \frac{dz}{N(z)^4} \right)^{-1} <0.
\ea
As we stated earlier, we do not  consider a finite volume bulk ($\theta_L=\theta_R=-1$). Let us now analyse the alternatives.  We see immediately that there is a spin 1 ghost whenever $\theta_L=\theta_R=+1$.  There are two spin 0 modes, roughly corresponding to the average radion, and the difference. The latter is never a ghost, whereas the kinetic term for the average radion is  determined by the sign of $f(2H^2)/F'(H^2)$.  Recall that for a well behaved cosmology we require that $F'(H^2) \geq 0$~\cite{cgp}. For finite $F'(H^2)>0$, it follows that we have a radion ghost whenever $f(2H^2)<0$.   In section \ref{sec:spin2}, we found that the lightest  spin 2 mode contains an helicity-0 ghost in precisely the opposite regime, ie when $f(2H^2)>0$. This is exactly the sort of behaviour found on the self-accelerating branch of DGP: a well behaved spin 2 sector corresponds to a pathological radion, and vice versa~\cite{arethereghosts, moreonghosts, dgpspec, review} .

When  $\theta_L=-1,~\theta_R=+1$, the spin 1 mode, and the radion difference decouple completely. In contrast, the average radion typically remains in the spectrum, and we can draw similar conclusions regarding its stability as discussed in the previous paragraph for $\theta_L=\theta_R=+1$. However, there are a few exceptional cases.  $F'(H^2)=0$ and  $F'(H^2) \to \infty$  ultimately correspond to the "stealth" scenarios identified in~\cite{cgp},  where the brane is Minkowski as opposed to de Sitter.  The former is the conformal  or strong coupling limit  whereas the latter is the decoupling limit.  The ghost is absent in both cases. Naively, the case $f(2H^2)=0$ would also appear to be ghost free, since the kinetic term for the radion vanishes.  Actually, this conclusion is incorrect. $f(2H^2)=0$ corresponds to the case where the radion mixes with the spin 2 mode, rendering our analysis invalid. The mixing occurs because the lightest spin 2 mode has  the same mass as the spin 0 mode ($m_{ \tt {light}}^2=2H^2$). The two modes  cease to be orthogonal and a more careful analysis is required. This was done for the self-accelerating branch of the DGP model, where the ghost was shown to remain even when $f(2H^2)=0$~\cite{moreonghosts, dgpspec}, It is natural to expect the same behaviour here.

\section{Discussion} \lab{sec:disc}
In this paper we have considered the stability of de Sitter branes in the CGP model: an asymmetric generalisation of the DGP model.  These vacua include the analogue of the normal branch in DGP, as well as the self accelerating branch.  Whenever the background bulk has infinite volume, we have found, without exception, that linear perturbations about these vacua contain  ghosts. As for the self accelerating branch of DGP, there is always a ghost in either the spin 2 or spin 0 sector. If the spin 2 sector is well behaved, there is a  spin 0 ghost corresponding to the average radion. If the spin 0 sector is well behaved, the helicity-0 part of the lightest spin 2 mode is a ghost. A more careful analysis is required in the crossover region, when the two offending modes mix with one another. However, our experience from the self-accelerating branch of DGP would imply that the ghost remains even in this limit~\cite{moreonghosts, dgpspec}.   In the most pathological scenarios, there is yet another ghost corresponding to the antisymmetric spin 1 mode.

It is interesting to note that the only way to avoid ghosts in this model is to consider Minkowski branes. This was studied in detail in~\cite{cgp}, where certain interesting vacua were found to be ghost free. These vacua corresponded to the "stealth" models, and had the curious property of giving rise to power law acceleration in the presence of matter, before asymptoting to Minkowski space  at late times. Indeed, the stealth model realises the Cardassian cosmology of Freese and Lewis~\cite{cardassian}, as well as offering a possible resolution of the coincidence problem. Given these successes of these models, it is worth asking whether or not our analysis can shed any light on their consistency.

Of course, the stealth vacua do {\it not} include  de Sitter branes. In fact, the vacuum brane is Minkowski and is known to be ghost free, in contrast to the de Sitter branes considered here.  What we can say is that the introduction of a small brane cosmological constant introduces an instability in the stealth model. It is reasonable to extend this conclusion to any type of matter, at least for small $H$. It follows that the stealth model is unstable close to the asymptotically Minkowski limit. The question now remains: how dangerous is this instability?

 A ghost will terrorize the vacuum  if it couples to ordinary fields. The problem is that in a unitary theory, the ghost ought to carry negative energy, and can be produced in the vacuum along with ordinary fields without violating energy conservation.  In a Lorentz invariant theory, the ghost-non ghost production rate is divergent,  no matter how weak the coupling!  This occurs because one can always use Lorentz invariance to perform a boost on the 3-momentum cut-off in loop integrals.  However, a generic  Friedmann-Robertson-Walker brane automatically breaks Lorentz invariance, so the stealth model does not necessarily suffer from this catastrophic instability (for a related discussion, see~\cite{izumi}).  If the ghost only couples weakly to other fields, the ghost-non ghost production rate gets suppressed.

  The stealth model contains a decoupling scenario where the would be ghost decouples from the spectrum as $H \to 0$. This corresponds to the case where we have $k_L=0, ~\theta_L=-1$ and $k_R>0, ~\theta_R=+1$, so the cosmological dynamics is governed  by the following
 \begin{equation}
	\lab{eq:Friedmann}
	\rho = F(H^2) = 6 M_4^2 H^2 - 6 M_R^3 \left( \sqrt{H^2+k_R^2}-k_R \right) + 6 M_L^3H.
\end{equation}
For small $H$, it is easy enough to check that $f(2H^2) \sim -2M_R^2 k_R<0$, from which we conclude that there is a radion ghost. We know from~\cite{cgp} that the radion decouples in the Minkowski limit, so it must be weakly coupled at small $H$.  Given that the radion feeds into the brane bending mode, we can see this explicitly by considering the coupling of the brane bending mode to matter (see equation (\ref{mattercoupling})). The coupling strength is given by $1/F'(H^2) \sim H/3M_L^3$, which does indeed go to zero as $H \to 0$. We conclude that this particular stealth model will barely be affected by the ghost at small $H$, owing to the weakness of the coupling. At larger values of $H$, our  de Sitter brane analysis suggests that the ghost coupling becomes significant, but we cannot be sure that these results apply to a general FRW brane.

~\\
{\large \bf Acknowledgements}\\
We would like to thank Ruth Gregory, Christos Charmousis  and Takahiro Tanaka for useful discussions.
KK was supported by ERC, RCUK and STFC. AP was funded by a Royal Society University Research Fellowship.
FPS was supported by
``Funda\c{c}\~{a}o para a Ci\^{e}ncia e a Tecnologia (Portugal)",
with the fellowship's reference number: SFRH/BD/27249/2006.

\end{document}